\documentclass{Interspeech2024}

\interspeechcameraready

\title{Who Finds This Voice Attractive?\\A Large-Scale Experiment Using In-the-Wild Data}

\name[affiliation={1}]{Hitoshi}{Suda}
\name[affiliation={2,\dagger}]{Aya}{Watanabe}
\name[affiliation={2,3}]{Shinnosuke}{Takamichi}

\address{
  $^1$National Institute of Advanced Industrial Science and Technology (AIST), Japan\\
  $^2$The University of Tokyo, Japan\quad$^3$Keio University, Japan}
\email{suda.h@aist.go.jp, aya.watanabe@alumni.u-tokyo.ac.jp, shinnosuke\_takamichi@ipc.i.u-tokyo.ac.jp}


\usepackage{booktabs}
\usepackage{multicol}
\usepackage{multirow}
\usepackage{algpseudocode}
\usepackage{algorithm}
\usepackage{siunitx}

\makeatletter
\let\MYcaption\@makecaption
\makeatother
\usepackage{subcaption}
\makeatletter
\let\@makecaption\MYcaption
\makeatother
\begin{document}

\maketitle
\gdef\thefootnote{\fnsymbol{footnote}}\footnotetext[2]{Presently with NTT Corporation, Japan}\gdef\thefootnote{\arabic{footnote}}
 
\begin{abstract}
This paper introduces CocoNut-Humoresque, an open-source large-scale speech likability corpus that includes speech segments and their per-listener likability scores.
Evaluating voice likability is essential to designing preferable voices for speech systems, such as dialogue or announcement systems.
In this study, we let 885 listeners rate 1800 speech segments of a wide range of speakers regarding their likability.
When constructing the corpus, we also collected the multiple speaker attributes: genders, ages, and favorite YouTube videos.
Therefore, the corpus enables the large-scale statistical analysis of voice likability regarding both speaker and listener factors.
This paper describes the construction methodology and preliminary data analysis to reveal the gender and age biases in voice likability.
In addition, the relationship between the likability and two acoustic features, the fundamental frequencies and the x-vectors of given utterances, is also investigated.
\end{abstract}

\section{Introduction}
With the advancement of speech synthesis, synthesized voices are widely used in various situations, such as voice assistants and announcements at public facilities\cite{Wagner2019-yn}.
Improvements in voice design technology, such as voice cloning and controlling techniques, have led to the enhancement of text-to-speech (TTS) systems and voice conversion techniques capable of a wide range of voice qualities\cite{Arik2017-ks,Chou2019-rt,Casanova2022-xd}.
Such techniques have been expected to customize the synthesized voice according to their purposes.
For instance, if each user could customize the voice quality of dialogue systems, improving user experiences (UX) of a user-specific dialogue system would be possible, enabling more comfortable systems\cite{Yu2019-by,Tolmeijer2021-jb}.
Furthermore, in advertising and promoting some products, using attractive voices for the target customer segment can enhance the effectiveness of advertisements\cite{Burkhardt2007-sc}.
Based on these backgrounds, statistical analysis of attractive voices is necessary to design voices that listeners find attractive.
Considering this, what kind of voices are attractive? Which listeners feel the attractiveness of the voice?



Multiple studies focusing on voice likability have been reported.
In a study evaluating simple voices such as monophthongs, the impact of fundamental frequency ($F_0$) is investigated, showing that specific ranges of $F_0$ are attractive\cite{Borkowska2011-dl}.
The effects of the speech rate have also been studied by evaluating single-word voices\cite{Ferdenzi2013-nx}.
These studies have only assessed short-duration voices and have not evaluated the influence of speaking styles, including accent and prosody.
In a study investigating the likability of utterances by 800 speakers, a relationship between multiple acoustic features and likability was shown; however, a detailed discussion on the differences in ratings among listeners was not conducted\cite{Burkhardt2011-tu}.
In another study that examined evaluation methods, a sufficient discussion on the differences in evaluation among listeners also has not been achieved\cite{Gallardo2016-ej}.
As a whole, we focus on two main unresolved points.
First, these studies have mentioned the relationship with only basic acoustic features, and the relationship with features such as x-vectors\cite{Snyder2018-lk}, which can be utilized in TTS and voice conversion, has yet to be investigated.
Next, in these studies, the number of listeners is minimal, making it impossible to discuss the tendencies of likability among diverse listeners according to listener attributes such as gender and age.
Therefore, by conducting evaluation with a large number of listeners on a sufficient number of utterances, it becomes possible to discuss the tendencies of voice likability according to the attributes of the listeners.
In addition, the utterances need meaningful sentences to extract speaker representations and evaluate them to consider their accents and prosody.

In this study, we collected subjective likability ratings for the utterances of meaningful sentences and constructed the CocoNut-Humoresque corpus.
By employing a large-scale speech description corpus for evaluation, we gathered information on likability that considers not only primary attributes like ``male voice'' or ``high-pitched voice'' but also factors such as prosody, accent, speaking style, and the speaker's age.
Additionally, this corpus includes the listener attributes, enabling the analysis of voice likability for each listener and providing insights for voice design.
This paper describes the methodology for constructing this corpus and conducts a preliminary analysis to reveal differences in likability according to both the speaker and listener.
\ifinterspeechfinal
     The corpus is publicly available from \url{https://github.com/sarulab-speech/Coco-Nut}.
\else
     The corpus is publicly available from \texttt{[URL will be filled on the camera-ready version]}.
\fi

\section{CocoNut-Humoresque: A large-scale speech likability corpus} 
To investigate ``who likes which voice,'' we constructed the CocoNut-Humoresque corpus, which consists of short speech segments and their likability scores by multiple listeners whose IDs and attributes are also included.
To construct this corpus, we let 885 listeners rate 1800 speech segments in terms of their likability.

\subsection{Voice materials}
As voice materials for rating, we adopted Coco-Nut\cite{Watanabe2023-cf}, which is a publicly available speech corpus.
Coco-Nut was constructed from YouTube videos with several user comments about voice qualities. Therefore, the corpus includes speech segments with a wide range of attractive voice qualities from various video genres.
The corpus contains 7330 short Japanese speech segments, provided as monaural audio signals in \SI{44100}{Hz}.
In addition, this corpus includes descriptions of speech characteristics (e.g., speaker attributes, speaking style, and emotional information) for all the segments.
The descriptions were submitted by listeners through a crowdsourcing service.
The segments are split into the train, validation, and test sets.
Note that this corpus includes only voices rated as attractive by several users, and therefore, the corpus does not cover all kinds of voice qualities, including non-attractive ones.

\subsection{Corpus design}\label{sec:corpus-design}
First, we selected 1800 speech segments from Coco-Nut.
In our corpus, 1200 segments were selected from the train set, 300 were selected from the validation set, and 300 were selected from the test set.

\begin{figure}
\centering
\includegraphics[width=.83\linewidth]{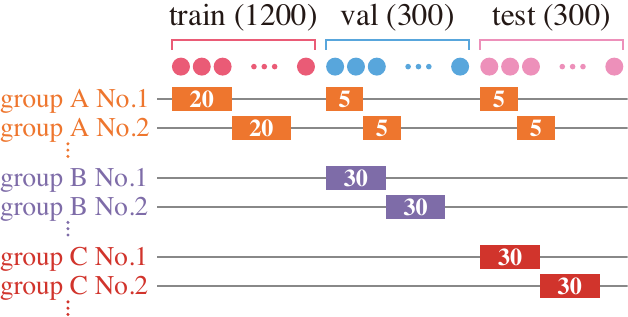}
\caption{Visualization of the contents of the 80 evaluation subsets in the corpus. The columns show the speech segments, and the rows show the subsets.}
\label{fig:subsets}
\end{figure}

Second, we constructed 80 evaluation subsets with 30 segments.
The first 60 subsets (group A) contain 20 segments, five segments, and five segments from the train, validation, and test sets, respectively.
The following ten subsets (group B) contain 30 segments selected from the validation set.
The last ten subsets (group C) contain 30 segments selected from the test set.
Hence, each segment from the validation or test set is evaluated by listeners for group A and group B or C.
Figure~\ref{fig:subsets} shows the visualization of the included segments in the subsets.
Each listener evaluated one subset selected randomly.
We equally allocate the listeners to 80 subsets to evaluate the subsets as well-balanced.
With the help of this subset construction method, the validation and test sets can be evaluated in both closed-listener and open-listener conditions when several systems (e.g., voice quality suggestion systems) are constructed.

\begin{algorithm}[tbp]
\caption{Create a subset $S$ from the whole segments.}\label{alg:subset}
\begin{algorithmic}[1]
\Require Subset size $n$, set of the whole segments $D$, and the x-vector extractor $v$.
\State $S \gets \emptyset$
\State $x \sim U(D)$
\Comment $x$ is randomly selected from $D$.
\While{$ |S| + 1 \leq n$}
    \State $S \gets S \cup \lbrace x\rbrace$
    \State $x \gets \operatorname{argmax}_{x \in D \backslash S} \sum_{y \in S} \lVert v(x)-v(y)\rVert^2$
\EndWhile
\State \Return $S$
\end{algorithmic}
\end{algorithm}

To keep the diversity of the speakers in each subset, we selected the segments to maximize the diversity of speaker embeddings\cite{Seki2024-ur}.
Algorithm~\ref{alg:subset} shows a pseudo-code of the algorithm for the subset construction.
A segment is randomly selected from the whole segments first, and then the most unlike segment from all segments in the subset is chosen repeatedly.
As speaker embeddings, we used WavLM-based x-vectors\cite{Snyder2018-lk,Chen2022-hq} extracted with \texttt{microsoft/wavlm-base-plus-sv}\footnote{\url{https://huggingface.co/microsoft/wavlm-base-plus-sv}}.
As a result, based on the descriptions in Coco-Nut, 1151 segments are male utterances, and 570 are female.
The gender of the remaining 79 utterances cannot be derived from the descriptions.
Note that this gender information is based on annotated descriptions in Coco-Nut, and the actual speakers' gender may differ from the perceived gender.
Hence, in constructing this corpus, rather than considering the balance of perceived genders, we selected the segments only based on the diversity of x-vectors.

In this annotation process, we asked each listener about his/her gender, age, and three favorite YouTube videos.
As for age, the listeners answered from the 10s, 20s, 30s, 40s, 50s, and over 59.
As for gender and age, the listeners are allowed to answer as N/A.
We also asked the listeners about the URLs of their favorite YouTube videos to reveal the relationship between their preference for videos and voice qualities.

The listeners scored the likability of voice qualities on a 6-point scale: 1---dislike entirely, 2---dislike, 3---slightly dislike, 4---slightly like, 5---like, or 6---completely like.
To avoid a score concentration on a neutral score, we did not adopt a 5-point or 7-point scale.
We asked the listeners to ignore the linguistic content and evaluate only the voice quality of the utterances.

We collected the listeners via an online crowdsourcing service.
The listeners can understand Japanese, the Coco-Nut corpus's primary language.
Each listener earned about 0.8 US dollars for his/her participation.
The listeners were not allowed to join this evaluation twice or more.

\subsection{Listener attributes}

\begin{table}[t]
\caption{The number of listeners in each age and gender. ``N/A'' means that the listener chose ``no answer.''}
\label{tbl:participants}
\vspace*{-2pt}
\footnotesize
\centering
\begin{tabular}{l|rrrrrrr|r}
\toprule
  &   10s &   20s &   30s &   40s &   50s &   60-- &   N/A &   Total \\
\midrule
 M   &     7 &    58 &   137 &   189 &    93 &     41 &     1 &     526 \\
 F &     1 &    45 &    97 &   131 &    64 &     15 &     2 &     355 \\
 N/A    &     0 &     1 &     1 &     0 &     0 &      0 &     2 &       4 \\
\midrule
 Total  &     8 &   104 &   235 &   320 &   157 &     56 &     5 &     885 \\
\bottomrule
\end{tabular}

\end{table}
\begin{table}[t]
\caption{The top five categories of listeners' favorite YouTube videos. Since some listeners did not answer their favorite videos or answered in an invalid format, the total number is less than the desired number.}
\label{tbl:favorite-videos}
\vspace*{-2pt}
\footnotesize
\centering
\begin{tabular}{rlr}
\toprule
   ID & Category &   Count \\
\midrule
         24 &  Entertainment &   525 \\
         10 &  Music &   376 \\
         22 &  People \& Blogs &   361 \\
         20 &  Gaming &   249 \\
         26 &  Howto \& Style &   213 \\
\midrule
         & Total & 2377 \\
\bottomrule
\end{tabular}

\end{table}

The number of listeners was 885.
Since the number of subsets is 80, each subset was assigned at least 11 listeners.
Both male and female listeners were assigned to each of the subsets.
Table~\ref{tbl:participants} shows the number of listeners by gender and age.
About 59\% of listeners are male, and 40\% are female.
While the average age was about 40, diverse listeners participated in this evaluation.
Table~\ref{tbl:favorite-videos} shows the categories of listeners' favorite videos. 
While the categories are not balanced, the diversity of the favorite videos was ensured.

\section{Analysis 1: Gender and age biases} 
This section compares distributions of the mean opinion scores (MOSs) of likability among different genders or ages to investigate gender and age biases in voice likability.
In this section, a MOS is calculated for each segment, and then the distributions of the MOSs are compared.

\subsection{Gender biases}\label{sec:gender-bias}



\begin{figure}
\centering
\includegraphics[width=.7\linewidth]{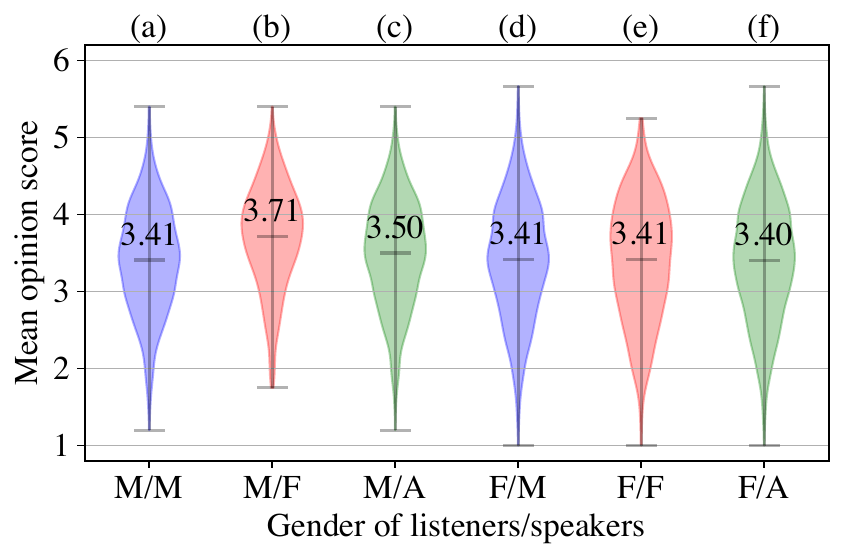}
\caption{Violin plots of the MOSs by genders of listeners and speakers. The left of each column's name denotes the listener's gender, and the right part denotes the speaker's gender. M, F, and A denote males, females, and whole data, respectively.}
\label{fig:mos_gender}
\end{figure}

In this section, gender biases in voice likability are investigated.
Distributions of MOSs are compared to clarify the effects of listeners' and speakers' genders.
Figure~\ref{fig:mos_gender} shows the violin plots of the results.
As described in Section \ref{sec:corpus-design}, the genders of speakers were estimated from the descriptions of the voice characteristics; therefore, in this paper, the genders of speakers mean the genders that the listeners sensed from the utterances.

First, the effects of the listener's gender are investigated.
With this investigation, we can observe the difference in scoring tendencies between males and females.
By comparing Figure~\ref{fig:mos_gender} (c) and (f), female listeners are confirmed to give lower scores to the voices than males.
This difference is significant since the $p$-value based on Welch's $t$ test is less than $7.0\times 10^{-5}$.
In addition, the $p$-value based on Bartlett's test was less than $1.9 \times 10^{-5}$, and therefore, the variance of scores by female listeners was also significantly more extensive than those by male listeners.

Next, the effects of the combination of listener's and speaker's genders are investigated.
By comparing Figure~\ref{fig:mos_gender} (a) and (b), for male listeners, the female utterances are more attractive than the male ones.
The $p$ value based on Welch's $t$ test is lower than $1.0 \times 10^{-5}$; therefore, this difference was significant.
On the other hand, by comparing Figure~\ref{fig:mos_gender} (d) and (e), for female listeners, no significant difference was observed in likability between the male and female utterances.

In conclusion, compared to female listeners, male listeners felt more attracted to female voices than male voices.
However, gender biases in speakers were not observed in female listeners.
That is, the female listeners rated the female utterances with lower scores than male listeners; hence, the female listeners gave lower scores in total.
In addition, the female listeners gave more distributed scores than the male listeners.
Therefore, as a whole, female listeners feel the equal likableness of male and female voices, and they can rate the voices on a broader range of scores.

\subsection{Age biases}

The effects of the listener's age are also investigated to reveal the differences in scoring tendencies among them.
In this investigation, the listeners were split into three groups: under 30, 30--49, and 50 or more.
In addition, only voices in validation and test sets were counted because these samples have more listeners than those in train sets.
By virtue of the corpus design, each utterance has at least 22 listeners.
A MOS was calculated for each utterance and age group.
If any age group has no listeners in an utterance, the utterance is eliminated.

\begin{figure}
\centering
\includegraphics[width=.7\linewidth]{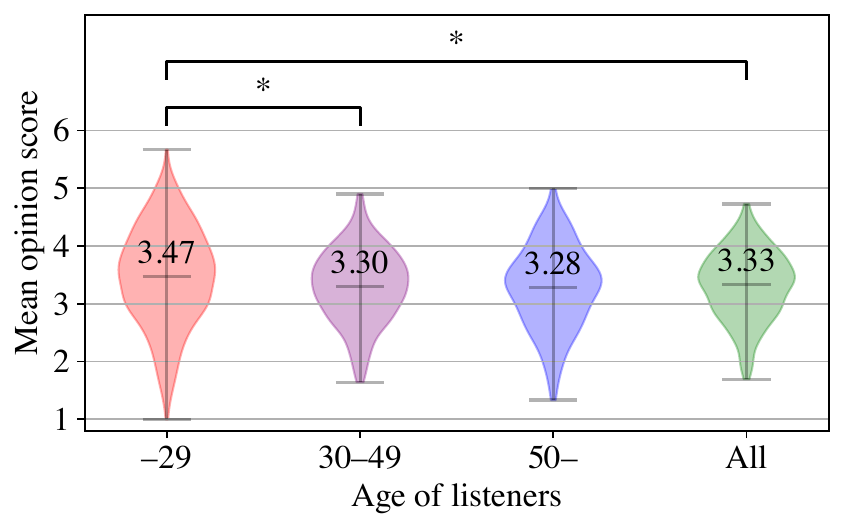}
\caption{Violin plots of the MOSs by the listener's age. Asterisks denote significant differences with $p<0.01$.}
\label{fig:mos_listener_age_la_sa}
\end{figure}

Figure~\ref{fig:mos_listener_age_la_sa} shows the violin plots of MOSs by age groups.
The younger listeners gave significantly higher scores to the voices since the $p$-value between the under-30 and the whole group was less than 0.01 based on Welch's $t$ test.
The variance in MOSs by younger listeners is also significantly more extensive than that of the whole group.

\begin{figure}
\centering
\includegraphics[width=.7\linewidth]{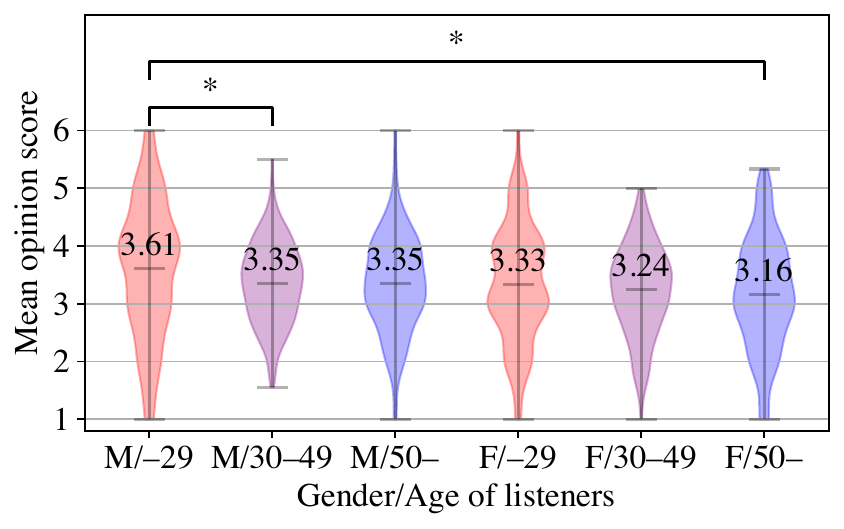}
\caption{Violin plots of the MOSs by the listener's gender and age. Asterisks denote significant differences with $p<0.01$.}
\label{fig:mos_listener_age}
\end{figure}

In addition, the effects of the combination of the listener's gender and age are investigated.
Figure~\ref{fig:mos_listener_age} shows the violin plots.
Based on the results, the discussed age biases are observed in both genders.
Additionally, it can be concluded that the younger male listeners gave high likability scores, and the older female listeners gave low scores since the $p$-value is less than $1.0\times 10^{-5}$ based on Welch's $t$ test.

\section{Analysis 2: Sample-by-sample analysis} 

This section investigates the voice likability sample by sample to reveal other likability factors than simple speaker attributes.

\subsection{Likable voices only for males or females}

\begin{figure}
\centering
\includegraphics[width=.9\linewidth]{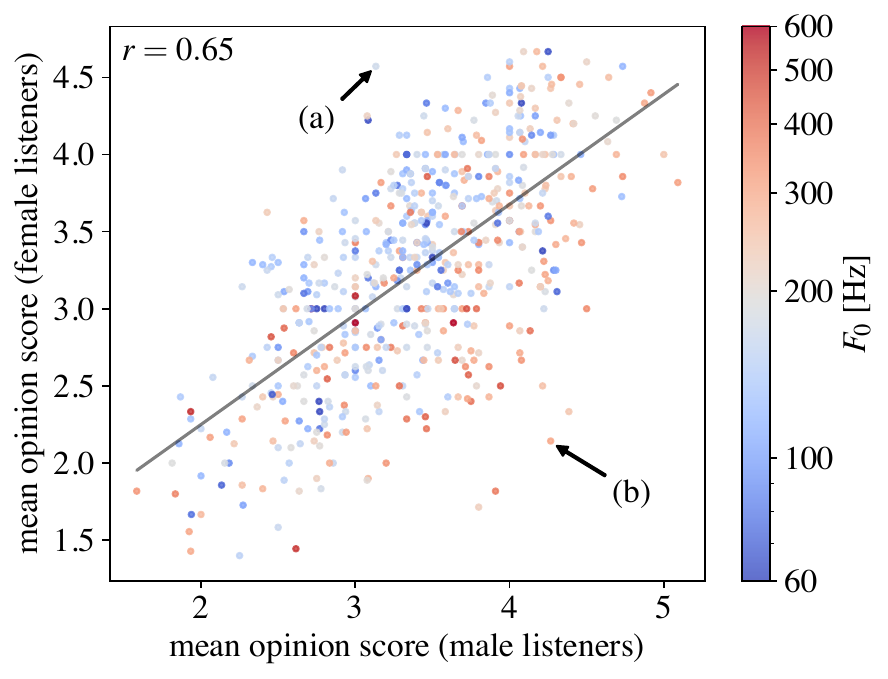}
\caption{Relationship between scores by male and female listeners. The samples in the upper left part are likable, especially for female listeners, and vice versa. The points (a) and (b) represent the samples with the most divided opinions between male and female listeners. The $F_0$ means were calculated using Crepe's full model\cite{Kim2018-ij}.}
\label{fig:mmos-fmos}
\end{figure}

To reveal which voices are likable for males but not for females or vice versa, the MOSs given by male and female listeners are compared.
In this investigation, utterances in the train set were not counted because of the few listeners.
Figure~\ref{fig:mmos-fmos} shows a chart of this relationship.
While a strong correlation between the male and female scores is observed, several samples appealed differently to male and female listeners.
In Figure~\ref{fig:mmos-fmos}, the utterances with the most notable differences between male and female scores are pointed out as points (a) and (b).
The utterance (a), a young male's acted speech, numbered 7230 in Coco-Nut, is likable, notably for female listeners.
On the other hand, the utterance (b), a young girl's relaxed speech, numbered 6979 in Coco-Nut, is likable, notably for male listeners.
As for utterance (b), while several male listeners rated it 6---completely like, several female listeners rated it 1---dislike entirely.
Hence, there exist voices that are greatly preferred differently by men and women.

As indicated in Figure~\ref{fig:mmos-fmos}, the pitch of the voices is one of the critical factors in the differences in preferences.
This result can be connected with the gender biases discussed in Section~\ref{sec:gender-bias}.
However, as the figure shows, some utterances with low mean $F_0$ are preferable, especially for men, and vice versa.
Therefore, the gender or the mean $F_0$ of a given utterance is not the only preference factor.
In addition, the trend that women feel more attracted to male utterances with low $F_0$ was not confirmed, while some papers point out it\cite{Feinberg2005-hy, Riding2006-qq}.
Hence, the attractiveness of the male voices to females can be affected not only by $F_0$ but also by other acoustic factors.

\subsection{Relationship between likability and x-vectors}

\begin{figure}
\centering
\begin{minipage}{0.99\linewidth}
\centering
\includegraphics[width=.73\linewidth]{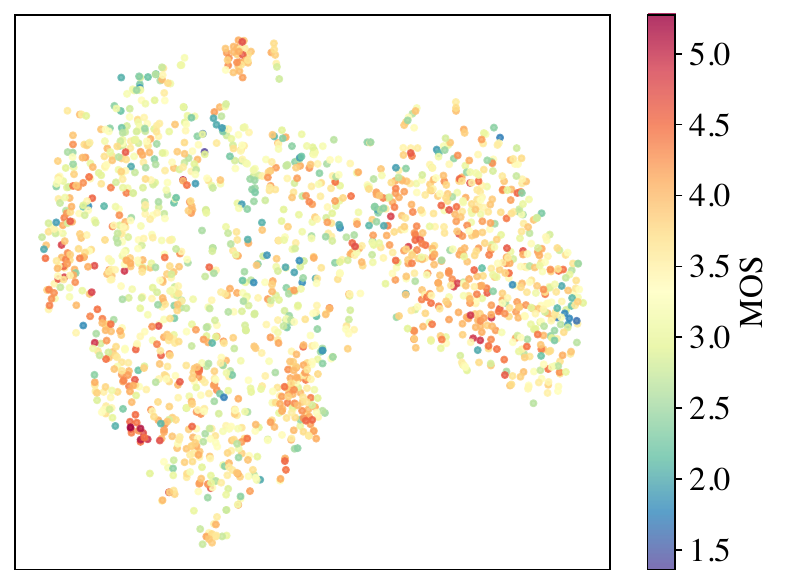}\vspace*{-5pt}
\subcaption{MOS.}
\label{fig:tsne_mean}
\end{minipage}
\begin{minipage}{0.99\linewidth}
\centering
\includegraphics[width=.73\linewidth]{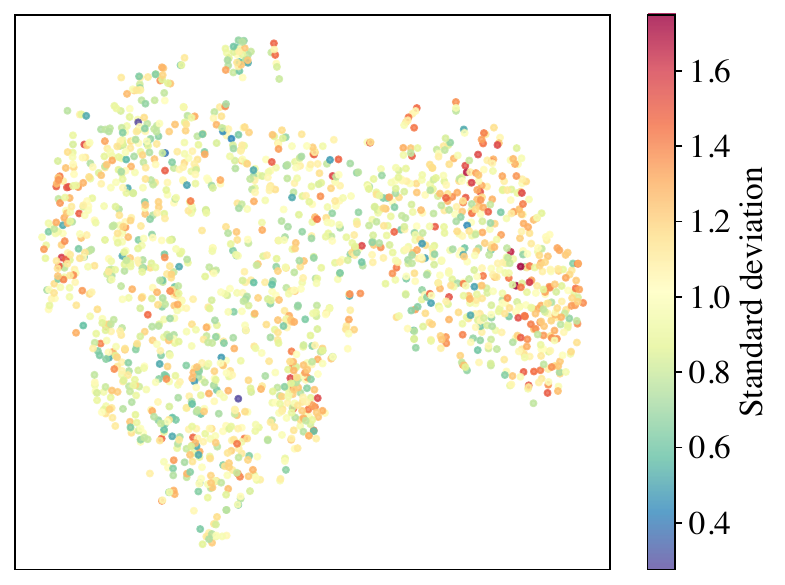}\vspace*{-5pt}
\subcaption{Standard deviation.}
\label{fig:tsne_stddev}
\end{minipage}
\caption{A $t$-SNE-based visualization of x-vectors colored by MOSs or standard deviations of the scores. In these figures, the left cluster contains male utterances, and the right cluster contains female ones.}
\label{fig:tsne}
\end{figure}

In this section, the relationship between x-vectors and likability scores is investigated.
By this investigation, we can determine whether the voice likability can be estimated from the speaker embeddings.
As x-vectors, WavLM-based x-vectors were extracted with \texttt{microsoft/wavlm-base-plus-sv}.
By applying $t$-SNE and reducing dimensions, the x-vectors are visualized.
Figure~\ref{fig:tsne} shows the results.
In this figure, the x-vectors formed clusters based on the speaker genders; the male utterances form the left cluster, and the female ones form the right.
As the figure indicates, a likability trend exists in an x-vector space for both the mean and standard deviation.
Voices in some areas are likable without high standard deviation, that is, without separating opinions.
On the other hand, voices in some areas have separate opinions among listeners, such as the right end of the female cluster.

These results indicate that the average voice likability or the variability in voice likability can be estimated using acoustic features such as $F_0$ and x-vectors of given utterances.
The voices with high likability or high variability do not concentrate on one area; nonetheless, there are a few subspaces for such utterances.

In addition, by comparing Figure~\ref{fig:tsne_mean} and Figure~\ref{fig:tsne_stddev}, some voices are confirmed to have nearly unified opinions regardless of their average likability scores.
Therefore, we can derive a voice with a specific stable score for some applications, such as a voice design system.

\section{Conclusions}

This paper introduces the CocoNut-Humoresque corpus, a large-scale speech likability corpus, to investigate voice likability.
The number of evaluated utterances is 1800, and the number of participating listeners is 885.
The evaluated utterances are natural Japanese utterances, and therefore, the voice likability can be investigated in in-the-wild situations.
By preliminary corpus analysis, gender biases were observed in both listeners and speakers, and age biases of listeners were also observed.
In short, male listeners gave higher scores to the female utterances than the male ones, and young male listeners gave higher scores than old female listeners.
In addition, by sample-by-sample analysis, the likability trends were observed according to $F_0$ and x-vectors of utterances.
However, voice likability is confirmed to be affected not only by these simple factors but also by other complicated factors.

\ifinterspeechfinal
\section{Acknowledgements}
     This work was supported by JSPS KAKENHI Grant Number 23K20017, 21H04900, 22H03639, and 23H03418, and JST FOREST JPMJFR226V.
     This paper is based on results obtained from a project, JPNP20006, commissioned by the New Energy and Industrial Technology Development Organization (NEDO).
\else
\fi

\bibliographystyle{IEEEtran}
\bibliography{mybib}

\begin{thebibliography}{10}
\providecommand{\url}[1]{#1}
\csname url@samestyle\endcsname
\providecommand{\newblock}{\relax}
\providecommand{\bibinfo}[2]{#2}
\providecommand{\BIBentrySTDinterwordspacing}{\spaceskip=0pt\relax}
\providecommand{\BIBentryALTinterwordstretchfactor}{4}
\providecommand{\BIBentryALTinterwordspacing}{\spaceskip=\fontdimen2\font plus
\BIBentryALTinterwordstretchfactor\fontdimen3\font minus
  \fontdimen4\font\relax}
\providecommand{\BIBforeignlanguage}[2]{{%
\expandafter\ifx\csname l@#1\endcsname\relax
\typeout{** WARNING: IEEEtran.bst: No hyphenation pattern has been}%
\typeout{** loaded for the language `#1'. Using the pattern for}%
\typeout{** the default language instead.}%
\else
\language=\csname l@#1\endcsname
\fi
#2}}
\providecommand{\BIBdecl}{\relax}
\BIBdecl

\bibitem{Wagner2019-yn}
P.~Wagner, J.~Beskow, S.~Betz, J.~Edlund, J.~Gustafson, G.~Eje~Henter,
  S.~Le~Maguer, Z.~Malisz, {\'E}.~Sz{\'e}kely, C.~T{\aa}nnander, and
  J.~Vo{\ss}e, ``Speech synthesis evaluation --- state-of-the-art assessment
  and suggestion for a novel research program,'' in \emph{Proc. 10th {ISCA}
  Workshop on Speech Synthesis ({SSW} 10)}, Sep. 2019.

\bibitem{Arik2017-ks}
S.~{\"O}. Ar{\i}k, G.~Diamos, A.~Gibiansky, J.~Miller, K.~Peng, W.~Ping,
  J.~Raiman, and Y.~Zhou, ``Deep voice 2: multi-speaker neural
  text-to-speech,'' in \emph{Proc. 31st International Conference on Neural
  Information Processing Systems}, Dec. 2017, pp. 2966--2974.

\bibitem{Chou2019-rt}
J.-C. Chou and H.-Y. Lee, ``One-shot voice conversion by separating speaker and
  content representations with instance normalization,'' in \emph{Proc.
  Interspeech 2019}, Sep. 2019.

\bibitem{Casanova2022-xd}
E.~Casanova, J.~Weber, C.~D. Shulby, A.~C. Junior, E.~G{\"o}lge, and M.~A.
  Ponti, ``{YourTTS}: Towards zero-shot multi-speaker {TTS} and zero-shot voice
  conversion for everyone,'' in \emph{Proc. 39th International Conference on
  Machine Learning}, vol. 162, 2022, pp. 2709--2720.

\bibitem{Yu2019-by}
Q.~Yu, T.~Nguyen, S.~Prakkamakul, and N.~Salehi, ````{I} almost fell in love
  with a machine'': Speaking with computers affects self-disclosure,'' in
  \emph{Extended Abstracts of the 2019 {CHI} Conference on Human Factors in
  Computing Systems}, no. Paper LBW0255, May 2019, pp. 1--6.

\bibitem{Tolmeijer2021-jb}
S.~Tolmeijer, N.~Zierau, A.~Janson, J.~S. Wahdatehagh, J.~M.~M. Leimeister, and
  A.~Bernstein, ``Female by default? --- exploring the effect of voice
  assistant gender and pitch on trait and trust attribution,'' in
  \emph{Extended Abstracts of the 2021 {CHI} Conference on Human Factors in
  Computing Systems}, no. Article 455, May 2021, pp. 1--7.

\bibitem{Burkhardt2007-sc}
F.~Burkhardt, R.~Huber, and A.~Batliner, ``Application of speaker
  classification in human machine dialog systems,'' in \emph{Speaker
  classification I: Fundamentals, features, and methods}, C.~M{\"u}ller, Ed.,
  Berlin, Heidelberg, 2007, pp. 174--179.

\bibitem{Borkowska2011-dl}
B.~Borkowska and B.~Pawlowski, ``Female voice frequency in the context of
  dominance and attractiveness perception,'' \emph{Animal behaviour}, vol.~82,
  no.~1, pp. 55--59, Jul. 2011.

\bibitem{Ferdenzi2013-nx}
C.~Ferdenzi, S.~Patel, I.~Mehu-Blantar, M.~Khidasheli, D.~Sander, and
  S.~Delplanque, ``\BIBforeignlanguage{en}{Voice attractiveness: Influence of
  stimulus duration and type},'' \emph{\BIBforeignlanguage{en}{Behavior
  research methods}}, vol.~45, no.~2, pp. 405--413, Jun. 2013.

\bibitem{Burkhardt2011-tu}
F.~Burkhardt, B.~Schuller, B.~Weiss, and F.~Weninger, ````{W}ould you buy a car
  from me?'' --- on the likability of telephone voices,'' in \emph{Proc.
  Interspeech 2011}, Aug. 2011.

\bibitem{Gallardo2016-ej}
L.~F. Gallardo, ``A paired-comparison listening test for collecting voice
  likability scores,'' in \emph{Proc. Speech Communication; 12. {ITG}
  Symposium}, Oct. 2016, pp. 1--5.

\bibitem{Snyder2018-lk}
D.~Snyder, D.~Garcia-Romero, G.~Sell, D.~Povey, and S.~Khudanpur, ``X-vectors:
  Robust {DNN} embeddings for speaker recognition,'' in \emph{Proc. 2018 {IEEE}
  International Conference on Acoustics, Speech and Signal Processing
  ({ICASSP})}, Apr. 2018, pp. 5329--5333.

\bibitem{Watanabe2023-cf}
A.~Watanabe, S.~Takamichi, Y.~Saito, W.~Nakata, D.~Xin, and H.~Saruwatari,
  ``{Coco-Nut}: Corpus of {J}apanese utterance and voice characteristics
  description for prompt-based control,'' in \emph{Proc. 2023 {IEEE} Automatic
  Speech Recognition and Understanding Workshop ({ASRU})}, Sep. 2023.

\bibitem{Seki2024-ur}
K.~Seki, S.~Takamichi, T.~Saeki, and H.~Saruwatari, ``Diversity-based core-set
  selection for text-to-speech with linguistic and acoustic features,'' in
  \emph{Proc. 2024 {IEEE} International Conference on Acoustics, Speech and
  Signal Processing ({ICASSP})}, Apr. 2024, pp. 12\,351--12\,355.

\bibitem{Chen2022-hq}
S.~Chen, C.~Wang, Z.~Chen, Y.~Wu, S.~Liu, Z.~Chen, J.~Li, N.~Kanda,
  T.~Yoshioka, X.~Xiao, J.~Wu, L.~Zhou, S.~Ren, Y.~Qian, Y.~Qian, J.~Wu,
  M.~Zeng, X.~Yu, and F.~Wei, ``{WavLM}: Large-scale self-supervised
  pre-training for full stack speech processing,'' \emph{IEEE Journal of
  Selected Topics in Signal Processing}, vol.~16, no.~6, pp. 1505--1518, Oct.
  2022.

\bibitem{Kim2018-ij}
J.~W. Kim, J.~Salamon, P.~Li, and J.~P. Bello, ``Crepe: A convolutional
  representation for pitch estimation,'' in \emph{Proc. 2018 {IEEE}
  International Conference on Acoustics, Speech and Signal Processing
  ({ICASSP})}, Apr. 2018, pp. 161--165.

\bibitem{Feinberg2005-hy}
D.~R. Feinberg, B.~C. Jones, A.~C. Little, D.~M. Burt, and D.~I. Perrett,
  ``Manipulations of fundamental and formant frequencies influence the
  attractiveness of human male voices,'' \emph{Animal behaviour}, vol.~69,
  no.~3, pp. 561--568, Mar. 2005.

\bibitem{Riding2006-qq}
D.~Riding, D.~Lonsdale, and B.~Brown, ``The effects of average fundamental
  frequency and variance of fundamental frequency on male vocal attractiveness
  to women,'' \emph{Journal of nonverbal behavior}, vol.~30, no.~2, pp. 55--61,
  Jun. 2006.

\end{thebibliography}

\end{document}